\documentclass[twocolumn,aps,prl,epsf,showpacs]{revtex4}
\usepackage{amsmath}
\usepackage{epsfig}
\usepackage{epsf}
\usepackage{array}

\begin{document}

\title{Nonlinear transport through two-terminal strongly-correlated heterostructures: A~dynamical-mean-field approach}
\author{Satoshi Okamoto}
\altaffiliation[Electronic address: ]{okapon@ornl.gov}
\affiliation{Materials Science and Technology Division, Oak Ridge National Laboratory, Oak Ridge, Tennessee 37831, USA}
\date{\today }

\begin{abstract}
The dynamical-mean-field method is applied to investigate the transport properties of heterostructures 
consisting of a strongly-correlated electron system connected to metallic leads. 
The spectral function inside the correlated region is sensitive to the change of the interaction strength and bias voltage. 
Because of this sensitivity, current vs voltage characteristics of such heterostructures are rather nonlinear 
regardless of the detail of the potential profile inside the correlated region. 
The electronic properties such as the double occupancy are also changed by the bias voltage.

\end{abstract}

\maketitle


Fabrication and characterization of 
heterostructures involving transition-metal oxides are one of the 
main topics of current materials science \cite{Ahn99,Izumi01,Gariglio02,Ohtomo02,Chkhalian06,Brinkman07,Reyren07}. 
In many transition-metal oxides, 
the electron-electron and/or electron-lattice interactions are comparable to or larger than the electron kinetic energy \cite{Imada98}. 
Therefore, a variety of exotic properties, such as 
high-$T_c$ superconductivity in cuprates \cite{Bednorz96} and 
colossal magnetoresitance in manganites \cite{Chahara93}, occur. 
In order to realize ``oxide electronics'' devices utilizing such bulk properties \cite{Bozovic03,Sun96,Bowen03}, 
theoretical understanding of the transport properties of correlated heterostructures is 
of crucial importance.

In a closely related subject, 
quantum transport through interacting-electron systems has been one of the most active fields in nanoscience. 
A variety of correlation effects, such as Kondo effect and Coulomb blockade, 
have been intensively studied. 
However, most theoretical techniques developed in these areas deal with 
a small number of orbitals connected to reservoirs \cite{Meir93,Konig96,Rosch03,Oguri05,AlHassanieh06}. 
Therefore, theoretical techniques remain to be developed for the bulklike effects of correlation 
on transport through heterostructures, including  
the correlation-induced Mott transition and symmetry breaking.

In this Letter, I undertake theoretical investigation of the transport properties of 
strongly-correlated heterostructures. 
As a simple model for such heterostructures, 
I consider several layers of a strongly-correlated system connected to two metallic reservoirs. 
I focus on the steady-state nonequilibrium properties of such structures under finite bias voltage. 
For this purpose, I apply a layer extension of the dynamical-mean-field theory \cite{Georges96} (layer DMFT) 
combined with the Keldysh Green's function technique \cite{Keldysh65}, 
a method recently proposed by the author \cite{Okamoto07}. 
The layer DMFT consists of mapping the lattice problem to quantum impurity models subject to the self-consistency condition. 
To solve quantum impurity models, I apply the non-crossing approximation (NCA) \cite{Pruschke89,Pruschke93}. 
This impurity solver is far more accurate than the equation of motion decoupling scheme (EOM) used in the previous study \cite{Okamoto07}. 
This allows one to study the nonequilibrium steady-state properties of correlated heterostructures 
over a relatively wide range of parameters covering the bulk Mott metal-insulator transition. 
It is revealed that the current-voltage characteristics of correlated heterostructures are rather nonlinear 
regardless of the detail of the potential profile. 
This originates from the close interplay between carrier injections and strong-correlation effects, 
while the electronic properties, such as spectral functions and double occupancy, depend on the potential. 
In some cases, applied bias voltage produces gapped spectral functions. 
This behavior differentiates the correlated heterostructures and other small systems such as the quantum dots.

\begin{figure}[tbp]
\includegraphics[width=0.6\columnwidth,clip]{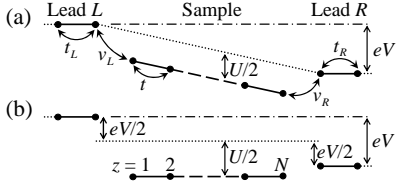}
\caption{Schematic view of the model heterostructure. 
(a) Linear potential profile for a sample with weak screening and (b) flat potential for strong screening. 
}
\label{fig:model}
\end{figure}

First, I outline the formalism of the present DMFT scheme (for more detail see Ref.~\onlinecite{Okamoto07}). 
I consider electrons moving on a cubic lattice with discrete translational invariance in the $xy$ plane. 
Each site is labeled by $\vec r = (\vec r_\parallel,z)$. 
A Hubbard-type interaction $U$ is introduced at a number $N$ of layers (sample $S$) located from $z=1$ to $N$, and 
noninteracting leads are located at $z \le 0$ (lead $L$) and $z \ge N+1$ (lead $R$). 
I consider the nearest-neighbor transfer $t$ ($t_\alpha$) of electrons in the sample (lead $\alpha$), 
the hybridization $v_\alpha$ between the sample and lead $\alpha$, 
and the layer-dependent potential $\varepsilon (z)$ (see Fig.~\ref{fig:model}). 
Thus, the Hamiltonian for this system is written as 
$H=H_S + \sum_{\alpha=L,R} (H_\alpha + H_{S-\alpha})$ with  
\begin{eqnarray}
H_S \!=\! - t \!\!\! \sum_{\langle \vec r, \vec r' \rangle, \sigma} \!\!\! \bigl( c_{\vec r \sigma}^\dag c_{\vec r' \sigma} + {\rm H.c.} \bigr)
+ U \sum_{\vec r} n_{\vec r \uparrow} n_{\vec r \downarrow} 
+ \sum_{\vec r, \sigma} \varepsilon (z) n_{\vec r \sigma} , \nonumber \\
%
H_\alpha \!=\! - t_\alpha \!\!\! \sum_{\langle \vec r, \vec r' \rangle, \sigma} \!\!\!
\bigl( c_{\vec r \sigma}^\dag c_{\vec r' \sigma} + {\rm H.c.} \bigr) 
+ \sum_{\vec r, \sigma} \varepsilon (z) n_{\vec r \sigma}, 
\hspace{6em}
\nonumber \\ 
%
H_{S-L(R)} \!=\! - v_{L(R)} \sum_{\vec r_\parallel, \sigma} 
\Bigl\{ c_{\vec r_\parallel (\vec r_\parallel + N \hat z) \, \sigma}^\dag c_{\vec r_\parallel + \hat z (\vec r_\parallel + N \hat z) \, \sigma} 
+ {\rm H.c.} \Bigr\}. 
\nonumber
\end{eqnarray}
Here, 
$c_{\vec r \sigma}$ is an electron annihilation operator at position $\vec r$ with spin $\sigma$, and 
$n_{\vec r \sigma} = c_{\vec r \sigma}^\dag c_{\vec r \sigma}$. 
The position $\vec r$ in each term is constrained as explained above and $\hat z = (0,0,1)$.

The hybridization $v_\alpha$ and the interaction $U$ are turned on adiabatically \cite{Caroli71,Datta}. 
The chemical potentials of two leads $\mu_L$ and $\mu_R$ 
and the site potentials $\varepsilon(z \le 0) = \varepsilon_L$ and $\varepsilon(z \ge N+1) = \varepsilon_R$ 
are assumed to be unchanged. 
In general, the potential profile in the sample should be computed self consistently by including long-range Coulomb interactions 
\cite{Cahay87}. 
Further, several layers of two leads should be considered as a part of the sample 
because the proximity effect would modify the electronic properties \cite{Freericks01,Okamoto04b}. 
I defer such self-consistent calculations for future work since it requires more parameters specific to the system under consideration. 
Instead, two extreme cases are considered as shown in Fig.~\ref{fig:model}. 
A realistic potential profile is expected to be between these two. 

After integrating out the lead degrees of freedom, one focuses on the sample 
in which the electron self-energy has two sources; electron correlations and the coupling with the leads. 
In the layer DMFT \cite{Potthoff99,Freericks01,Okamoto04b,Okamoto07}, 
the self-energy due to correlations is approximated to be diagonal in layer index $z$ and independent of in-plane momentum $k_\parallel$. 
Thus, the lattice self-energy is written as 
$
\Sigma_{z,z'}^\gamma \bigl(\vec k_\parallel, \omega \bigr) 
\Rightarrow \delta_{z,z'} \bigl\{ \Sigma_z^\gamma (\omega) 
+ v_L^2 g_L^\gamma \bigl(\vec k_\parallel, \omega \bigr) \delta_{z,1}
+ v_R^2 g_R^\gamma \bigl(\vec k_\parallel, \omega \bigr) \delta_{z,N} 
\bigr\}, 
$
where $\gamma =r$ and $K$ stand for the retarded and Keldysh components of the Green's function, respectively, 
and $g_\alpha^\gamma \bigl(\vec k_\parallel, \omega \bigr)$ the surface Green's function of lead $\alpha$.

The correlation part of the self-energy is computed by introducing a number $N$ of quantum impurity models 
subject to the self-consistency condition of DMFT; 
the impurity Green's function $G_{imp,z}^\gamma(\omega)$
and the local part of the lattice Green's function $G_{loc,z}^\gamma(\omega)$ are identical: 
\begin{eqnarray}
G_{imp,z}^\gamma (\omega) 
= G_{loc,z}^\gamma (\omega) \equiv \! \int \! \frac{d^2k_\parallel}{(2\pi)^2} G^\gamma_{zz} \bigl(\vec k_\parallel, \omega \bigr). 
\label{eq:Gloc}
\end{eqnarray}
The lattice Green's function matrix $\hat G^\gamma (\vec k_\parallel, \omega)$ is given by 
$
\hat G^r \bigl( \vec k_\parallel, \omega \bigr) = 
\bigl[ \omega+i0_+ - \hat H_S \bigl( \vec k_\parallel; U=0) 
- \hat \Sigma^r \bigl( \vec k_\parallel, \omega \bigr) \bigr]^{-1} , 
$
and
$
\hat G^K \bigl(\vec k_\parallel, \omega \bigr) =
\hat G^r \bigl(\vec k_\parallel, \omega \bigr) \hat \Sigma^K \bigl(\vec k_\parallel, \omega \bigr) 
\hat G^{r*} \bigl(\vec k_\parallel, \omega \bigr)$. 
The impurity model at layer $z$ is now 
characterized by the hybridization function $\Delta_z^\gamma (\omega)$ and 
the effective distribution function of electrons $f_{eff,z} (\omega)$. 
%
These are fixed by Eq.~(\ref{eq:Gloc}) as 
$
\Delta_z^r (\omega) = 
\omega - \varepsilon (z) - \Sigma_z^r (\omega)
- \bigl\{ G_{loc,z}^r (\omega) \bigr\}^{-1}
$
and 
$
G^K_{loc,z} (\omega) = 2 i \bigl\{ 1-2f_{eff,z} (\omega) \bigr\} 
{\rm Im} \, G_{imp,z}^r (\omega)
$ \cite{feff}.
 
In order to solve the impurity model by NCA \cite{Pruschke89,Pruschke93}, 
four kinds of auxiliary particles are introduced: 
bosonic $e (d)$ representing an empty (doubly occupied) state and 
fermionic $f_\sigma$ a single occupied state by an electron with spin $\sigma$ 
and a local constraint $e^\dag e + d^\dag d + \sum_\sigma f_\sigma^\dag f_\sigma = 1$. 
The local constraint is treated by introducing a complex chemical potential \cite{Bickers87}. 
The retarded, advanced and lesser Green's functions of the auxiliary particles 
are computed self-consistently to update the electron self-energy $\Sigma_z^\gamma(\omega)$ 
which will be used in the next iteration.  
After the self-consistency is obtained, the lattice Green's functions are used to compute physical quantities.

The EOM used in Ref.~\cite{Okamoto07} 
was found to underestimate the critical interaction $U_c$ for the metal-insulator transition; 
for the $N \rightarrow \infty$ limit of my model (3-dimentional Hubbard model), $U=10t$ gives an insulating solution. 
A more accurate exact diagonalization (ED) impurity solver \cite{Georges96} with 8-site cluster 
estimates $U_c \approx 16t$. 
Present NCA reproduces $U_c$ of ED within few percent.

In the following, I mainly use parameters $v_{L,R}=t$, $t_{L,R}=2.5t$. 
Numerical results do not depend on these parameters in a significant way 
for $|eV| < 6 t_{L,R}$ beyond which the finite band width of leads starts to contribute. 
I focus on half-filled case at $eV=0$ taking $\varepsilon_{L,R} = \mu_{L,R}$ with $T=0.1t$ 
and only consider paramagnetic states. 

\begin{figure}[tbp]
\includegraphics[width=0.75\columnwidth,clip]{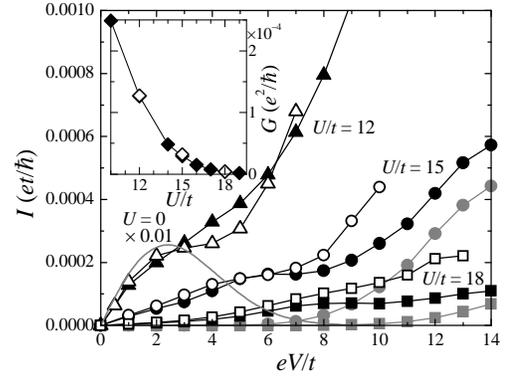}
\caption{Current-voltage characteristics of $N=6$ heterostructure with $U=12t$ (triangle), $15t$ (circle), $18t$ (square). 
Filled (open) symbols are the results for linear (flat) potential, and 
light line corresponds to the linear potential with $U=0$ multiplied by $0.01$. 
For comparison, EOM results are also shown for the linear potential with $U=15t$ and $18t$ as light symbols.
Inset: Linear conductance at $V=0$ as a function of $U$. 
Filled (open) symbols are for the linear (flat) potential.}
\label{fig:IV}
\end{figure}

Figure \ref{fig:IV}
plots the current vs bias voltage for $N=6$ heterostructure with several choices of on-site interaction: 
$U=12t (<U_c)$, $15t (\alt U_c)$, $18t (>U_c)$, and 0 as a reference. 
Although there is quantitative difference, 
two choices of potential profile give similar curves for all finite $U$'s. 
In particular, two curves overlap at small bias voltage where the transport is governed by (induced) quasiparticle band. 
In this region, clear crossover between the metallic and insulating regions can be seen in the linear conductance $G$ 
(inset of Fig.~\ref{fig:IV}). 
Compared with the result for $U =0$, 
the current and the conductance are reduced substantially corresponding to the reduction of the quasiparticle weight. 
The conductance decreases nearly linearly with increasing $U$ in this window and, above the critical value, becomes exponentially small. 
In the insulating region, $G$ decreases exponentially with $N$ 
because of the exponential decay of the induced quasiparticle weight. 
By increasing a bias voltage, carriers are injected and the quasiparticle weight grows 
causing an upturn in the $I$-$V$ curve for $U =18t$. 
%
At the intermediate bias, there appear ``plateaus'' for small $U$. 
In this region, chemical potentials of two leads touch the Hubbard bands. 
Thus, electrons in the less-developed quasiparticle bands suffer from strong scattering.

When the chemical potentials of the two leads enter the Hubbard bands further, 
larger spectral weight overcomes the effect of scattering. 
This causes further upturn in the current, and the two curves start to deviate. 
In this region, tunneling between neighboring Hubbard bands was found to carry the current for the linear potential \cite{Okamoto07}. 
As shown by light symbols, the EOM reproduces the position and magnitude of the current density well. 
For flat potential, carriers are more strongly injected into the Hubbard bands. 
Thus the tunneling picture no longer holds in this region.

For flat potential, one encounters a well-known problem of NCA, the breakdown of analycity 
(the imaginary part of the retarded self-energy becomes positive), 
at very large bias; $eV \agt 7t$ for $U=12t$, $eV \agt 10t$ for $U=15t$ and $eV \agt 13t$ for $U=18t$. 
This prevents one from further computation. 
Including the vertex correction for the self-energies of auxiliary particle is expected to remedy this problem. 
Even at $eV \sim 7t$ for $U=12t$, $eV \sim 10t$ for $U=15t$ and $eV \sim 13t$ for $U=18t$, 
some parts of ${\rm Im} \, \Sigma_z^r(\omega)$ are found to be positive. 
However, coupling with leads generates an additional negative imaginary part of the self-energy 
producing causal solutions.

\begin{figure}[bp]
\includegraphics[width=0.8\columnwidth,clip]{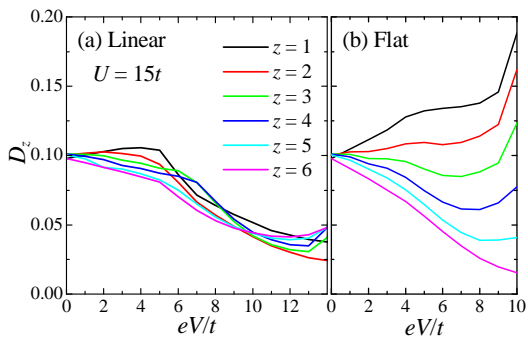}
\caption{(Color online) Position-dependent double occupancy for $N=6$ heterostructure with $U=15t$ as a function of bias voltage. 
(a) Linear potential and (b) flat potential. 
}
\label{fig:double}
\end{figure}

Although two choices of potential give rather similar $I$-$V$ curves, 
the electronic properties were found to differ significantly at the intermediate-to-large bias region. 
Figure~\ref{fig:double} plots the position-dependent double occupancy 
$D_z = \langle n_{z\uparrow} n_{z\downarrow} \rangle = \langle d_z^\dag d_z \rangle$ with $U=15t \alt U_c$ 
for (a) linear potential and (b) flat potential. 
At $eV \alt 5t$ (roughly the distance between the upper and lower Hubbard bands), 
the double occupancy gradually increases at layers near lead $L$ and decreases near lead $R$ 
for both choices of potential. 
For the flat potential, this trend continues until $eV \sim 7t$ beyond which
the electron injection into the upper Hubbard band near lead $R$ becomes significant. 
On the other hand, for the linear potential, the double occupancy in all layers starts to decrease beyond $eV \sim 5t$ until $eV \sim 12t$. 
Similar phenomena are observed at a bulk Mott metal-insulator transition \cite{Georges96}.

\begin{figure}[bp]
\includegraphics[width=0.85\columnwidth,clip]{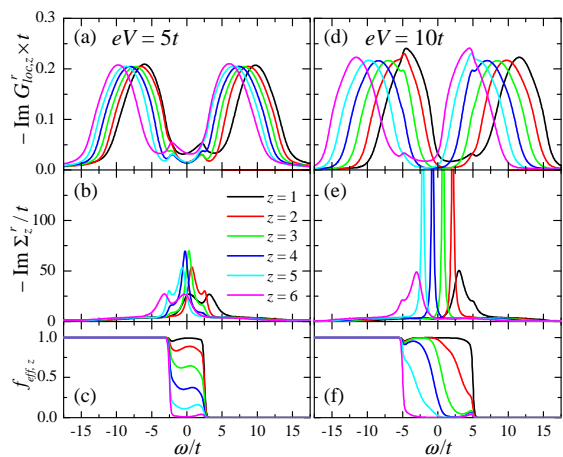} 
\caption{(Color online) Imaginary parts of the local Green's function $G_{loc,z}^r$ and 
the self-energy $\Sigma_z^r$, and effective distribution function $f_{eff,z}$ 
for $N=6$ heterojuction with $U=15t$ and linear potential profile. 
Applied bias voltage is (a--c) $eV=5t$ and (d--f) $eV=10t$. 
}
\label{fig:GSF_L}
\end{figure}

\begin{figure}[tbp]
\includegraphics[width=0.85\columnwidth,clip]{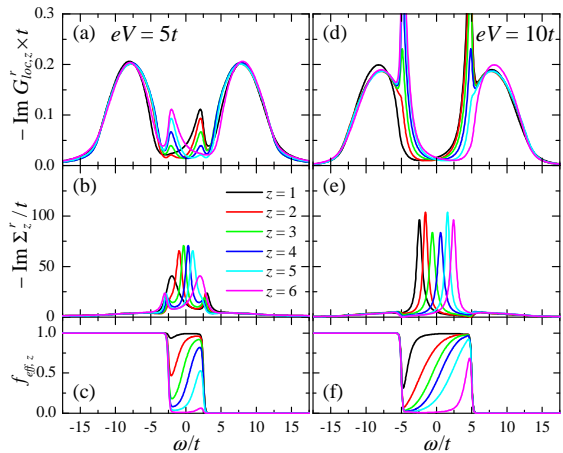}
\caption{(Color online) Same as in Fig.~\ref{fig:GSF_L} with flat potential. 
}
\label{fig:GSF_F}
\end{figure}

To further clarify the difference between the two potentials, 
I plot in Fig.~\ref{fig:GSF_L} 
the imaginary parts of the local Green's function 
and the retarded self-energy, 
as well as the effective distribution function 
for $N=6$ heterostructure with the linear potential.
At each bias voltage, one observes signals of ``quasiparticle'' peaks at the Fermi levels of two leads, $\omega = \pm eV/2$. 
Strikingly, the spectral weight between the upper and lower Hubbard bands is suppressed for layers $2 \le z \le 5$ 
when the bias voltage is increased. 
This also accompanies the divergence of ${\rm Im} \Sigma_z^r$. 
Although, some amount of carriers are already injected and the electric current is flowing inside the Hubbard bands, 
the suppression of the spectral function has stronger effect on the double occupancy (compare ${\rm Im} G_{loc,z}^r$ and $f_{eff,z}$). 
Close inspection of the distribution function reveals that such a behavior is due to 
the increase in the ``effective temperature'' that electrons feel. 
As shown in Fig.~\ref{fig:GSF_L} (f), $f_{eff,z}$ varies over a rather wide range of $\omega$ of the order of the applied voltage. 
The gapped spectral functions explain why EOM and NCA give similar $I$-$V$ curves at large bias. 
Similar enhancement of the self-energies is also observed for flat potential 
[compare Figs.~\ref{fig:GSF_F}(b) and \ref{fig:GSF_F}(e)]. 
But the spectral functions do not have a well defined gap. 
Instead, one observes very strong quasiparticle peaks at $\omega = \pm eV/2$ because a large number of carriers are injected. 
Further, the effective temperature looks even reversed at $|\omega| < \pm eV/2$ 
[see Figs.~\ref{fig:GSF_F}(bc) and \ref{fig:GSF_F}(f)].

So far, the simple potential profiles are considered, linear and flat, and  
nonlinear current-voltage characteristics were found to be rather insensitive to the profile. 
Thus, similar behavior is expected even for a realistic potential profile determined self-consistently. 
On the other hand, the electronic properties were dependent on the potential profile. 
Thus, in the realistic situation, smooth transition between the two solutions is expected to occur as follows: 
the potential gradient will be suppressed near the interface because of the screening by the injected carriers, 
leading to the sharp quasiparticle peaks. 
The finite slope is expected to remain deep inside the sample where the injected electrons and holes are nearly balanced, 
thus the gapped spectral functions would be realized at the large bias voltage. 

Aside from the complexity associated with self-consistent potential calculations, 
several extensions of the present work are desirable. 
Magnetic symmetry breaking, a characteristic of correlated systems, can be included in the present formalism. 
Possible melting of magnetic ordering by an applied voltage and 
its effect on the transport properties is an interesting problem. 
The present procedure can be also applied to Anderson and Kondo type models, 
and multiorbital and multisite models. 
Including 
$d$-wave superconducting correlations for a multisite model \cite{Maier00} 
is an interesting application for the transport properties of junctions involving high $T_c$ cuprates. 
Since the present formalism is simple, combining it with the density functional theory would not be difficult. 

To summarize, 
I investigated the transport properties of heterostructures consisting 
of a strongly correlated system connected to metallic leads by using the layer DMFT. 
The current vs bias voltage characteristics of such heterostructures are found to be nonlinear. 
This originates from the sensitivity of the single particle spectral function inside the correlated region against the bias voltage. 
These effects may also become crucial for interpreting or predicting 
phenomena in which a correlated system is necessarily driven out of equilibrium.

The author thanks J. E. Han, B. K. Nikoli{\'c}, S. Onoda, and Z. Y. Zhang for their valuable discussions.  
This work was supported by the Division of Materials Sciences and Engineering, the U.S. Department of Energy.  


\begin{thebibliography}{99}
\bibitem{feff}Further, I employ the ansatz 
$\frac{\Sigma_z^K}{{\rm Im} \Sigma_z^r} = \frac{\Delta_z^K}{{\rm Im} \Delta_z^r} = 2i(1 -2 f_{eff,z})$ \cite{Okamoto07}, 
which becomes exact in equilibrium. 
In the present formalism, it guarantees the current conservation. 
Possible improvements by going beyond the ansatz are the broadening of $f_{eff,z}$ and the suppression of quasiparticle peaks 
at large bias. 
In general NCA can determine $\Sigma_z^K$ and $\Delta_z^K$ independently, 
but reliable results are unavailable so far because of substantial numerical error.
\end{thebibliography}
\end{document}